\documentclass[twocolumn,showpacs,preprintnumbers,amsmath]{revtex4-1}
\usepackage{dsfont,epsfig,amsmath,amssymb,rotating,graphicx}
\begin{document}
\title{ Non-nequilibrium model on Apollonian networks}
\bigskip

\author{ F. W. S. Lima}
\affiliation{Dietrich Stauffer Computational Physics Lab, Departamento de F\'{\i}sica, 
Universidade Federal do Piau\'{\i}, 64049-550, Teresina - PI, Brazil}.
\email{fwslima@gmail.com}
\author{Andr\'e A. Moreira}
\author{Asc\^anio D. Ara\'ujo}
\affiliation{Departamento de F\'{\i}sica, Universidade Federal do
Cear\'a, Campus do Pici, 60451-970 Fortaleza, Cear\'a, Brazil.}
\email{auto@fisica.ufc.br}
\email{ascanio@fisica.ufc.br}

\begin{abstract}
We investigate the Majority-Vote Model with two states ($-1,+1$) and a noise $q$ on Apollonian networks. The main result found here is the presence of the phase transition as a function of the noise parameter $q$. We also studies de effect of redirecting a fraction $p$
of the links of the network.
By means of  Monte Carlo simulations, we obtained the exponent ratio $\gamma/\nu$, $\beta/\nu$, and $1/\nu$  for several values of rewiring probability $p$. The critical noise was determined $q_{c}$ and $U^{*}$ also was calculated. The effective dimensionality of the system was observed to be independent on $p$, and the value $D_{eff} \approx1.0$ is observed for these networks.
Previous results on the Ising model in Apollonian Networks have reported no presence of a phase transition. Therefore, the results present here demonstrate that the Majority-Vote Model belongs to a different universality class as the equilibrium Ising Model on Apollonian Network. 
 \end{abstract}

\maketitle

\section{Introduction}
The Ising model \cite{a3,onsager} is commonly used  as a
benchmark to test and improve new algorithms and methods for computer
simulation in models of Statistical Mechanics. For instance, Monte
Carlo methods such as Metropolis \cite{me}, Swendsen-Wang \cite{s-w},
Wang-Landau \cite{w-l}, Single histogram \cite{f-s} and
Broad histogram \cite{pmco} have all been used to calculate the critical
exponents of this model. The Ising model has also been employed to
study social behavior and many of these models and others can be found out in
\cite{book}.

G. Grinstein {\it et al.} \cite{g} have argued that non-equilibrium
stochastic spin systems on regular square lattices (SL) with up-down
symmetry, fall into the same universality class of the equilibrium
Ising model. The correspondence was observed for several
models that do not obey detailed balance and on other regular lattices
\cite{C,J,M,mario,lima01,lima02}. The majority-vote model with two
states (MV2) is a non-equilibrium model proposed by M.J. Oliveira in
$1992$ which does not obey detailed balance. This model follows a stochastic 
dynamics with local rules and with up-down symmetry, and on a regular lattice shows a second-order
phase transition with critical exponents $\beta$, $\gamma$, $\nu$
consistent \cite{mario,a1} with those of the equilibrium Ising model
\cite{a3}.
However, Lima {\it et al.} \cite{lima0} have studied MV2 on Voronoi-Delaunay random
lattices with periodic boundary conditions, and they obtained exponents
different from those obtained on regular lattices, in disagreement
with the conjecture suggested by G. Grinstein {\it et al.} \cite{g}.
 
 Simulations on both {\it undirected} and {\it directed} scale-free
networks \cite{newman,sanchez,ba1,alex,sumour,sumourss,lima}, random
graphs \cite{erdo,er2} and social networks \cite{er,er1,DS}, have
attracted interest of researchers from various areas. These complex
networks have been studied extensively by Lima {\it et al.} in the context
of discrete models (MV2, Ising, and Potts model)
\cite{lima1,lima2,lima3,lima4,lima5,lima6}. Recently,
the equilibrium Ising model was studied on a class of hierarchical
scale-free networks, namely the Apollonian Networks (ANs)
\cite{andrade1,andrade2}, and it was shown that,  on these networks,
no phase transition is observed for these models.

In the present work, we study the MV2 model on normal and redirected
ANs. By means of numerical simulations we found that MV2 model in ANs network 
displays a clear second-order phase transition. This demonstrates that, on these networks, 
MV2 and the Ising model do not fall in the same universality class, therefore contradicting
 Grinstein hypothesis \cite{g}. The remainder of our paper,
is organized as follows. In section 2, we present our model and some
details about the Monte Carlo simulation as well as the calculations
performed in the evolution of the physical quantities. In section 3,
we do an analysis over the simulations performed and discuss the
obtained results. Finally in section 4, we present our conclusions and
final remarks.

\section{Model and simulation } 

Our network is ANs type composed of $N=3+(3^{n}-1)/2$ nodes, where $n$
is the generation number \cite{Andrade_05}. To introduce a level of disorder we redirect a faction $p$
of the of the links. The redirecting results in a directed network, preserving the out-going node of the redirected link
but changing the incoming node. In the limit of $p=0$  we have the Apollonian Networks,
while in the limit $p=1$ we have something similar to random networks \cite{erdo}. Note however,
that the number of outgoing links of each node is preserved, therefore, even in the limit $p=1$
the network still have hubs that that are the most influent nodes. For $p=0$ we have the
standard  Apollonian Networks. These networks display a scale free degree distribution and
a hierarchical structure. The critical properties of percolation and Potts models on these networks have been
investigated \cite{andrade2,auto} and its was shown that in the thermodynamic limit there is no phase transition,
with the ordered phase prevailing for any finite temperature.

On the MV2 model, the system dynamics is as follows. 
Initially, we assign a spin variable
$\sigma$ with values $\pm 1$ at each node of the network. At each step
we try to spin flip a node. The flip is  accepted
with probability 
\begin{equation}
w_{i}(\sigma)=\frac{1}{2}\biggl[ 1-(1-2q)\sigma_{i}S\biggl(\sum_{\delta
=1}^{k_{i}}\sigma_{i+\delta}\biggl)\biggl],
\label{eq_1}
\end{equation}
where $S(x)$ is the sign $\pm 1$ of $x$ if $x\neq0$, $S(x)=0$ if
$x=0$. To calculate $w_i$ our sum runs over the number $k$ of nearest
neighbors of $i$-th spin.  Eq.~(\ref{eq_1}) means that
with probability $(1-q)$ the spin will adopt the same state as the majority
of its neighbors. 
Here, the control parameter $0\le q\le 1$
plays a role similar to the temperature in equilibrium systems, the smaller
$q$ greater the probability of parallel aligning with the local majority.

We performed Monte Carlo simulation on the ANs with various systems of
size $N=3,283$; $~9,844$; $~29,527$; $~88,576$, 
and $265,723$.  We wait
$1\times 10^5$ Monte Carlo steps (MCS) in order to reach the steady
state, and then the time averages are estimated over the next $2\times
10^5$ MCS.  One MCS is accomplished after all the $N$ spins are
investigated whether they flip or not.  We carried out $N_{sample}=1,000$
to $10,000$ independent simulation runs for each lattice and for a
given set of parameters $(q,N)$.  

To study the critical behavior of the model we define the variable
$m=\sum^{N}_{i=1}\sigma_{i}/N$. In particular, we are interested in
the magnetization $M$, susceptibility $\chi$ and the reduced
fourth-order cumulant $U_{4}$
\begin{equation}
M(q)=\biggl[ \langle|m|\rangle\biggl]_{av},
\label{eq_2}
\end{equation}
\begin{equation}
\chi(q)= N\biggl[\left(\langle m^2\rangle-\langle m \rangle^2\right)\biggl]_{av},
\label{eq_3}
\end{equation}
\begin{equation}
U_{4}(q)= 1-\biggl[\langle m^{4}\rangle/\left( 3\langle m^2 \rangle^2 \right)\biggl]_{av},
\label{eq_4}
\end{equation}
where $\langle\cdots\rangle$ stands for a thermodynamics average.
The results are averaged over the $50$ (av) ANs independent simulations. 
These quantities are functions of the noise parameter $q$ and obey the finite-size
scaling relations
\begin{equation}
M=N^{-\beta/\nu}f_m(x),
\label{eq_5}
\end{equation}
\begin{equation}
\chi=N^{\gamma/\nu}f_\chi(x),
\label{eq_6}
\end{equation}
\begin{equation}
\frac{dU}{dq}=N^{1/\nu}f_U(x),
\label{eq_7}
\end{equation}
where $\nu$, $\beta$, and $\gamma$ are the usual critical 
exponents, $f_{i}(x)$ are the finite size scaling functions with
\begin{equation}
x=(q-q_c)N^{1/\nu}
\label{eq_8}
\end{equation}
being the scaling variable.  From this scaling relations we obtained
the exponents $\beta/\nu$ and $\gamma/\nu$, respectively. Moreover, the
value of $q^*$ for which $\chi$ has a maximum is expected to scale
with the system size as
\begin{equation}
q^*=q_c+bN^{-1/\nu}
\label{eq_9}
\end{equation}
where $b\approx 1$. Therefore, these relations may be used to obtain
the exponent $1/\nu$.
  
\begin{figure}[t]
\includegraphics[width=7cm]{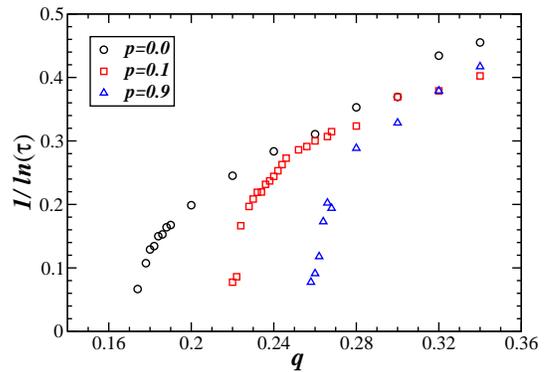}
\caption { Reciprocal logarithm of the relaxation times $\tau$ on directed
  Apollonian Networks versus $q$ for different redirecting probability ({\small{$\bigcirc$}}) $p=0.0$, ($\square$) $p=0.1$, ($\triangle$) $p=0.9$.}
\label{fig_1}
\end{figure} 

\section{ Results and discussion}

In order to test, if there is a phase transition in MV2 models we
measured the relaxation time $\tau$ as a funcion of the noise
parameter $q$. We start the system 
with all spins up, a number of spins equal to $7,174,456$ ($G15$).
We determine the time $\tau$ after which the magnetization has flipped
its sign for the first time, and then take the median value of nine
samples. As one can see in Fig.~\ref{fig_1}, the relaxation time goes to
infinity at some positive $q$ value, indicating a second order phase
transition. On contrast, the Ising model on directed Barabasi-Albert \cite{ba1} networks has no
phase transition and agrees with the modified Arrhenius law for
relaxation time \cite{lima-edina}.
\begin{figure}[t]
\includegraphics[width=7cm]{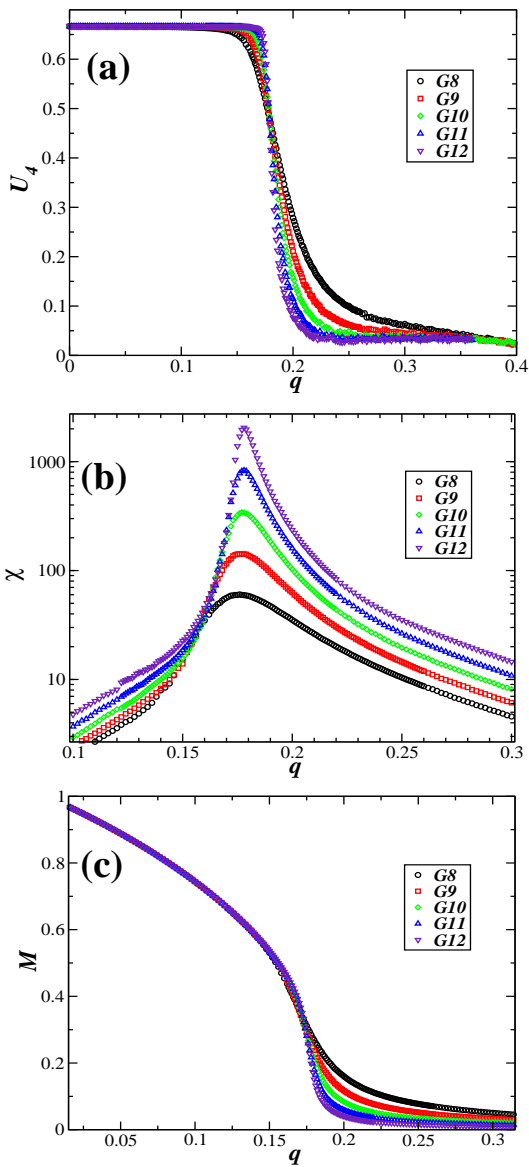}
\caption{We show the state functions $U_{4}$ (a) , $\chi$ (b), and  $M$ (c) studied here as a function of the noise $q$ for the Apolloanian Network (redirecting probability $p=0$). The results are presented for different generations of the Appolonian Network,({\small{$\bigcirc$}}) $G8$, ($\square$) $G9$,    ({\Large{$\diamond$}}) $G10$, ($\triangle$) $G11$, ({\large$\triangledown$}) $G12$. The Binder cumulant clearly presents a the phase-transition, with the critical noise being obtained by the point where the curves intercept each other, $q_c \approx 0.17$ . The signature of the phase transition is also observed in the curves for the susceptibility and magnetization.}
\label{fig_2}
\end{figure}

\begin{figure}[t]
\includegraphics[width=7cm]{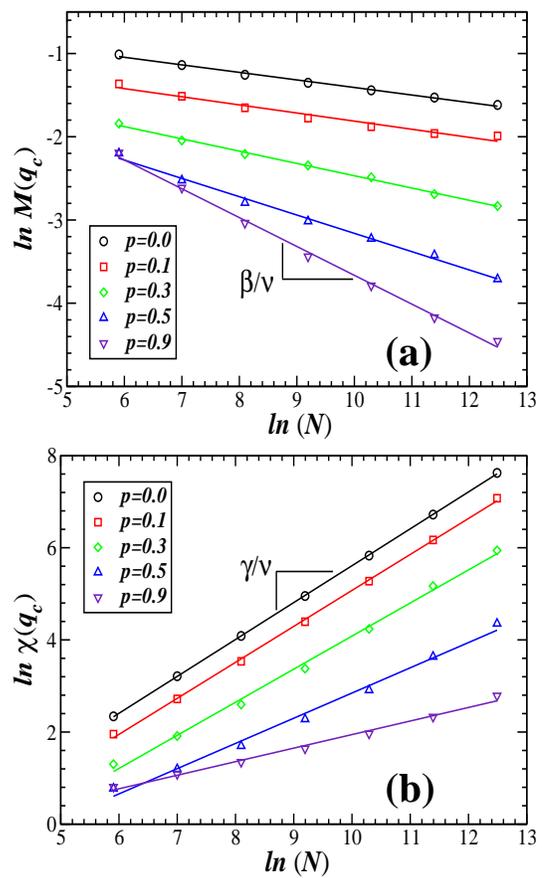}
\caption{The Log-log plot of the magnetization $M(q_{c})$ (a), susceptibility $\chi(q_{c})$ (b) as a function of the number of nodes $N$ on the Apollonian Networks. Each curve of magnetization and susceptibility were calculated at the critical point $q_{c}$ for each value of $p$ as shown in table~1. The different symbols correspond to following redirecting probabilities: ({\small{$\bigcirc$}}) $p=0.0$, ($\square$) $p=0.1$, ({\Large{$\diamond$}}) $p=0.3$, ($\triangle$) $p=0.5$ and ({\large$\triangledown$}) $p=0.9$. In both graphs (a) and (b) the solid lines represent the best linear fit for each value of $p$ and the slopes give $\beta/\nu$ and $\gamma /\nu$, respectively.}
\label{fig_3}
\end{figure}

\begin{figure}[t]
\includegraphics[width=7cm]{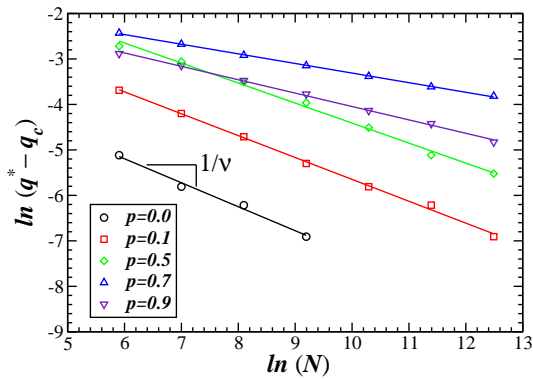}
\caption{Log-log plot of the displacement of the point of maximum of the susceptibility $(q^{*}-q_{c})$ agains the number of nodes $N$ on the Apollonian Networks. The symbols represent the following values of $p$:  ({\small{$\bigcirc$}}) $p=0.0$, ($\square$) $p=0.1$, ({\Large{$\diamond$}}) $p=0.5$, ($\triangle$) $p=0.7$ and ({\large$\triangledown$}) $p=0.9$. The solid lines indicate the best linear regression and from the straigh lines slopes we obtain the exponent $1/\nu$ according to Eq.~(\ref{eq_9}).}
\label{fig_4}
\end{figure}

\begin{figure}[t!]
\includegraphics*[width=8.5cm]{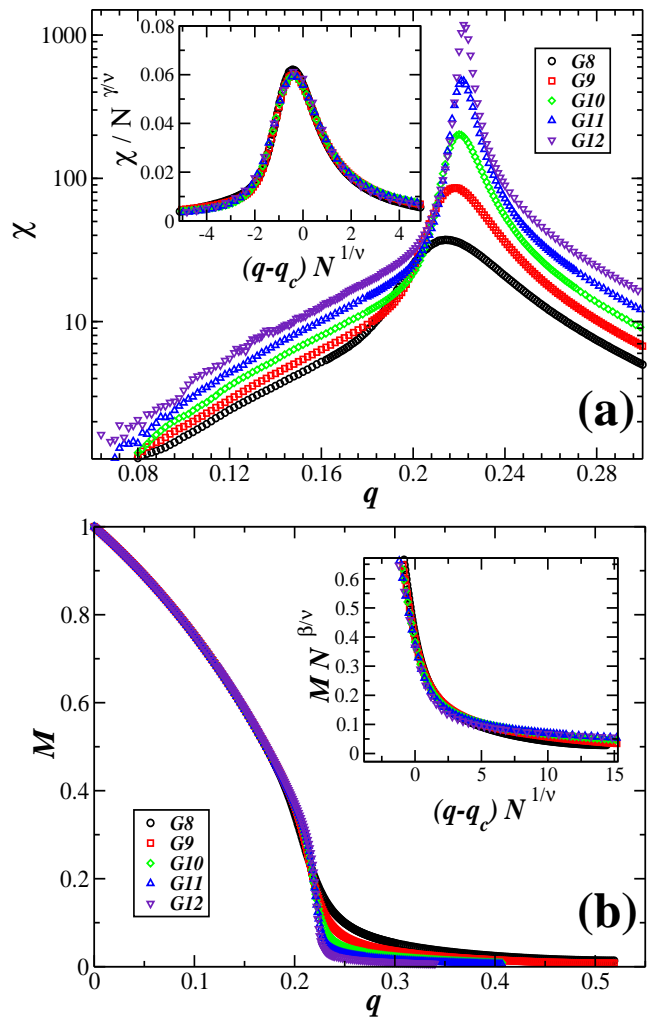}
\caption{We show the state functions  $\chi$ (a) and  $M$ (b) as a function of the noise $q$ for directed Apolloanian Networks (redirecting probability $p=0.1$). The results are presented for different generations of the Appolonian Network,({\small{$\bigcirc$}}) $G8$, ($\square$) $G9$,    ({\Large{$\diamond$}}) $G10$, ($\triangle$) $G11$, ({\large$\triangledown$}) $G12$. In the insets we show the rescaled data, collapsed using the critical exponents obtained from table~1.}
\label{fig_5}
\end{figure}

In Fig.~\ref{fig_2}, we show the dependence of the magnetization $M$, Binder's
cumulant $U_{4}$ and susceptibility $\chi$ on the noise parameter $q$,
obtained from simulations on ANs with $N=3,283$; $~9,844$; $~29,527$; $~88,576$, 
and $265,723$ sites and with $n=8$, $9$, $10$, $11$, and $12$ generation
($G8$, $G9$, $G10$, $G11$,~and~$G12$). The shapes of magnetization 
curves atest the presence of a second-order phase transition in the
system.  The critical noise parameter $q_c$ is estimated as
the point where the curves for different system sizes $N$ intercept
each other \cite{binder}. The obtained values for $q_c$ can be seen in table ~1.
The critical exponents $\beta/\nu$ and $\gamma/\nu$ can be obtained by investigating the scaling at criticality
of the magnetization and susceptibility, respectively. As shown in Fig.~\ref{fig_3}, both
quantities scale as power laws with the controlling exponents depending on the redirecting probability.
We use Eq.~(\ref{eq_9}) to obtain the exponent $1/\nu$ as shown in Fig.~\ref{fig_4}. In table~1 we summarize 
all critical exponents for each values of $p$. In Fig.~\ref{fig_5} we plot respectively the susceptibility and magnetization as a function of $q$ for systems with redirecting probability $p=0.1$ and different generations of the Apollonian Networks. Using the scaling exponents and Eqs.~(\ref{eq_2}) and (\ref{eq_3}) we produced the collapsed data shown in the insets, which confirm the accuracy of the exponents.  

We can also compute the effective dimensionalities of the system, defined as
$D_{eff}=2\beta/\nu + \gamma/\nu$÷  \cite{Privman}. For all values of $p$ we obtain $D_{eff}\approx 1$, as seen in table~1. This
behavior  has been previously observed for MV2 and
MV3 on various Scale-free networks and on Erd\"os-R\'enyi random
graphs~\cite{erdo}. 

\begin{table}[t]
\begin{center}
\begin{tabular}{|c||c|c|c|c|c|c|}
\hline
$p$ & $q_{c}$ & $1/\nu$ &  $\beta/\nu$ & $\gamma/\nu$ & Deff\\
\hline
$0.0$ & $0.178(3)$ & $0.53(4)$ & $0.091(3)$ & $0.80(2)$ & $0.98(3) $ \\
\hline
$0.1$ & $0.223(5)$ & $0.48(2)$ & $0.097(5)$ & $0.79(3)$ & $0.98(3) $ \\
\hline 
$0.2$ & $0.249(3)$ & $0.66(3),$ & $0.112(3)$ & $0.80(3)$ & $ 1.02(2)$\\
\hline 
$0.3$ & $0.271(5)$ & $0.61(8)$ & $0.148(5)$ & $0.72(5)$ & $1.02(3) $  \\
\hline
$0.4$ & $0.284(5)$ & $0.71(7)$ & $0.130(3)$ & $0.69(4)$  & $ 0.95(6)$ \\
\hline
$0.5$ & $0.296(4)$ & $0.44(5)$ & $0.220(8)$ & $0.55(3)$  & $0.99(5) $ \\
\hline
$0.6$ & $0.313(3)$ & $0.23(3)$ & $0.343(4)$ & $0.32(2)$  & $1.01(3) $ \\
\hline
$0.7$ & $0.311(5)$ & $0.21(5)$ & $0.374(5)$ & $0.25(3)$  & $ 1.01(4)$ \\
\hline
$0.8$ & $0.290(5)$ & $0.27(3)$ & $0.36(2)$ & $0.28(3)$  & $1.00(2) $ \\
\hline
$0.9$ & $0.2629(3)$ & $0.29(6)$ &  $0.347(9)$ & $0.29(2)$  & $ 0.98(5)$ \\
\hline
\end{tabular}
\end{center}
\caption{The critical noise $q_{c}$, and the critical exponents, 
for  ANs with probability $p$. Error bars are statistical only.} 
\end{table}

%

\section{Conclusion}
We presented results for the non-equilibrium MV2 on ANs.  On this
network, the non-equilibrium MV2 shows a well defined second-order
phase transition. On other hand,  it has been previously shown that the equilibrium Ising model 
does not present a phase transition in these networks~\cite{andrade1,andrade2}. Therefore our results demonstrate
that MV2 model on ANs belongs to another universality class, in
disagreement with the conjecture of Grinstein et al.
\cite{g}. The source of this distinction is due to the different behavior of noise 
in each of these models. In the Ising model, the probability of switching a highly connected
spin against the local majority is smaller than a less connected one; since the energy variation is larger for a more connected spin.
In the MV2 model, the probability of a spin switching against the local majority is always given by $q$, independent
on the number of neighbors of this spin. Interestingly, the effective dimensionality of the system, defined as
$D_{eff}=2\beta/\nu + \gamma/\nu$,  is always a value close 1.0 independent of the rewiring probability $p$, as seen in table~1. This
value for $D_{eff}$ has already been previously obtained for MV2 and
MV3 on various Scale-free networks and on Erd\"os-R\'enyi random
graphs. 

\subsection{Acknowledgments}
F. W. S. Lima thanks Dietrich Stauffer for many suggestions
and fruitful discussions during the development of this work. We thank
CNPq and FUNCAP for financial support. This work also was supported
the system SGI Altix 1350 the computational park CENAPAD.UNICAMP-USP,
SP-BRAZIL.

\end{document}